\documentclass{moriond}

\usepackage{float}
\usepackage[utf8]{inputenc}
\usepackage{amsfonts,amsmath,amssymb}
\usepackage[english]{babel}
\usepackage{array}
\usepackage{color}
\usepackage{bm}

\newcommand{\D}{\mathrm{d}}
\newcommand{\WKB}{\mathrm{WKB}}
\newcommand{\dB}{\mathrm{dB}}
\newcommand{\TE}{\mathrm{TE}}
\newcommand{\TM}{\mathrm{TM}}
\newcommand{\tz}{\tilde{z}}
\newcommand{\tF}{\tilde{F}}
\newcommand{\tpsi}{\tilde{\psi}}

\newcommand{\bz}{\bm{z}}
\newcommand{\bF}{\bm{F}}
\newcommand{\bE}{\bm{E}}
\newcommand{\bV}{\bm{V}}

\newcommand{\cW}{\mathcal{W}}
\newcommand{\cR}{\mathcal{R}}

\newcommand{\Hb}{\bar{\mathrm{H}}}
\newcommand{\gb}{\bar{g}}

\newcommand{\Photo}{}

\begin{document}
\vspace*{4cm}
\title{CASIMIR EFFECT AND QUANTUM REFLECTION}

\author{ G. DUFOUR, R. GU\'EROUT, A. LAMBRECHT, S. REYNAUD }

\address{Laboratoire Kastler Brossel, UPMC-Sorbonne
Universit\'es, CNRS, ENS-PSL Research University, Coll\`ege de
France, Campus Jussieu, F-75252 Paris, France.}

\maketitle

\abstracts{
The GBAR experiment will time the free fall of cold antihydrogen atoms dropped onto an annihilation plate to test the universality of free fall on antimatter.
In this contribution, we study the quantum reflection of the anti-atom resulting from the Casimir-Polder attraction to the plate.
We evaluate the Casimir-Polder potential and the associated  quantum reflection amplitudes and find that reflection is enhanced for weaker potentials. A Liouville transformation of the Schrödinger equation is used to map the quantum reflection problem onto an equivalent problem of scattering on a barrier, leading to an intuitive understanding of the phenomenon. 
}

\section*{Introduction}

The universality of free fall is tested  with ever increasing precision both for macroscopic test masses\cite{Wagner2012} and atoms \cite{Fray2004,Schlippert2014}.
However, a precise direct measurement of the free fall of antimatter is still lacking. The current experimental bound on the gravitational acceleration $\gb$ of antihydrogen was obtained by the ALPHA collaboration \cite{TheALPHACollaboration2013}: $-65 g \leq \gb \leq +110 g$.
Several experiments built around the CERN Antiproton Decelerator and the new deceleration ring ELENA will attempt to reduce this bound to the percent level in the coming years \cite{Knecht2014,Hamilton2014}.
In particular, the GBAR experiment (Gravitational Behavior of Antihydrogen at Rest) will time the free fall of cold antihydrogen ($\Hb$) atoms \cite{Chardin2011,Indelicato2014}. 
Following the method proposed in \cite{Walz2004}, a cold $\Hb$ will be obtained by photo-detaching the excess positron of an $\Hb^+$ ion that has been previously cooled down to the lowest quantum state in a Paul trap. 
The neutral $\Hb$ then falls in the Earth's gravity field with an acceleration $\gb= Mg/m$, where $M$ is the gravitational mass of $\Hb$, $m$ its inertial mass and $g$ the local gravity field.
Annihilation of the anti-atom on a material plate placed at a height $h$ below the ion trap marks the end of the free fall. Assuming no initial velocity, the value of $\gb$ can be inferred from the time between the photo-detachment pulse and the detection signal: $T=\sqrt{2h/\gb}$.
A more detailed description of the quantum wavepacket's motion gives the arrival time distribution around this mean value \cite{Dufour2014shaper}.

The above program assumes that no force other than gravity is acting on the anti-atom during the free-fall. Yet Casimir and Polder have shown \cite{Casimir1946,Casimir1948} that neutral atoms in the vicinity of a material medium experience an attractive force because quantum fluctuations of the electromagnetic field couple the atomic induced dipole to induced dipoles in the medium.
Within experimental accuracy, the Casimir-Polder (CP) force does not modify the free fall time. However, despite its attractive nature, it causes quantum reflection of atoms at low energies \cite{Friedrich2002,Voronin2005pra,Voronin2005jpb}.

Such classically forbidden reflection from an attractive potential is a manifestation of the wave-like behavior of quantum matter and it occurs when the potential varies rapidly on the scale of the de Broglie wavelength \cite{Berry1972,Mody2001,Friedrich2002}.
Experiments have observed quantum reflection on the CP potential near liquid \textrm{He} \cite{Nayak1983,Berkhout1989,Yu1993} and solid surfaces \cite{Shimizu2001,Druzhinina2003,Pasquini2004}, as well as rough or micro/nanostructured surfaces \cite{Shimizu2002ridged,Pasquini2006,Zhao2010}.

Theory \cite{Mody2001,Judd2011} and experiments \cite{Shimizu2002ridged,Pasquini2006} have shown that the reflection probability increases when the energy is reduced, but also when the absolute magnitude of the potential is decreased. For example, it is larger for atoms falling onto a silica mirror than onto
silicon or metallic mirrors \cite{Dufour2013qrefl} and it is even larger for nanoporous silica \cite{Dufour2013porous} which couples extremely weakly to the electromagnetic field.
We give an intuitive explanation of this paradoxical effect by mapping the quantum reflection problem into a problem of
reflection on a repulsive barrier by means of a Liouville transformation of the Schrödinger equation. Liouville transformations are exact mappings of Schrödinger equations onto one another which have the remarkable property of preserving scattering amplitudes.

In this contribution, we will first sketch how the scattering approach to Casimir forces \cite{Lambrecht2006,Emig2007} can be used to accurately calculate the interaction of $\Hb$ with various types of surfaces. We then compute the associated quantum reflection, seen as the exchange between counter-propagating WKB waves. In section \ref{sec:liouville} we introduce Liouville transformations of the Schrödinger equation and show how a specific choice of coordinate maps scattering on an attractive well onto reflection from a repulsive wall. We finish by looking into the high reflection limit of low energies and weak potentials and discuss its relevance to high precision tests of the Equivalence Principle.

\section{The Casimir-Polder potential}
\label{sec:CP}

In the scattering approach to Casimir forces, the interaction energy between two arbitrary objects in vacuum is written in terms of the reflection operators $\cR_1$, $\cR_2$ for electromagnetic waves on each object  \cite{Lambrecht2006}. For distances $z$ between the objects below the thermal wavelength ($\sim 1\ \mu$m at 300~K), one can use the zero-temperature expression:
 \begin{equation}\label{Casimir}
    V(z)=\hbar \int_0^\infty \frac{\D\xi}{2 \pi} \text{Tr} \log\left( 1-\cR_1 e^{-\kappa z}\cR_2 e^{-\kappa z}\right)~.
  \end{equation} 
This formula is obtained after a Wick rotation to imaginary frequencies $\omega=i\xi$, which transforms the oscillating terms $e^{i k_z z}$ describing translation from one object to the other into decaying exponentials $e^{-\kappa z}$, $\kappa= \sqrt{\vec{k}_\perp^2+\xi^2/c^2}$. In these formulas, $\vec k = \vec{k}_\perp\pm k_z\vec e_z$ is the wavevector of the electromagnetic radiation. The trace runs over the transverse wavevector $\vec{k}_\perp$ and the polarizations $\TE,\TM$.

We now specialize to the case of an atom above a plane.
The reflection operator on the plane is diagonal in the plane wave basis where it is given by the Fresnel reflection amplitudes $\rho^\TE$, $\rho^\TM$. These reflection coefficients depend on the material properties of the medium through its relative dielectric function $\varepsilon(\omega)$.
We treat the atom in the dipolar approximation, so its reflection operator depends on the dynamic polarizability $\alpha(\omega)$, which is supposed to be the same as that of (ground state) hydrogen. Finally we neglect multiple reflections on the atom by expanding the logarithm in \eqref{Casimir} to first order and find  \cite{Messina2009,Dufour2013qrefl}:
\begin{equation}\label{Casimir-Polder}
V(z)=\frac{\hbar}{c^2} \int_0^\infty\!\!\!\!\!\! \D\xi\ \xi^2 \alpha(i\xi) \int \!\! \frac{\D^2\vec{k}_\perp}{(2\pi)^2} \frac{e^{-2 \kappa z}}{\kappa} \left[ \rho^{TE}-\left(1+\frac{2c^2k_\perp^2}{\xi^2}\right)\rho^{TM} \right]~.
\end{equation}
This formalism allows an easy inclusion of realistic optical response properties for the atom and for all types of mirrors. Those used in this work are detailed in Ref. \cite{Dufour2013qrefl}.

The typical wavelength $\lambda$ characterizing the optical response of the atom and plane defines a transition between two asymptotic behaviors of the CP potential. For a thick mirror:
\begin{equation}
\label{vlimits} V(z) \underset{z \ll \lambda}{\to} -\frac{C_3}{z^3}~,\qquad\qquad V(z)\underset{z \gg \lambda}{\to} -\frac{C_4}{z^4}~.
\end{equation}
The short distance limit is the well known van der Waals potential; whereas the large separation limit is referred to as the \emph{retarded} CP interaction since it takes into account the finiteness of the speed of light  \cite{Casimir1946,Casimir1948}.

\begin{figure}[h]
\centering
 \includegraphics[width=0.45\textwidth]{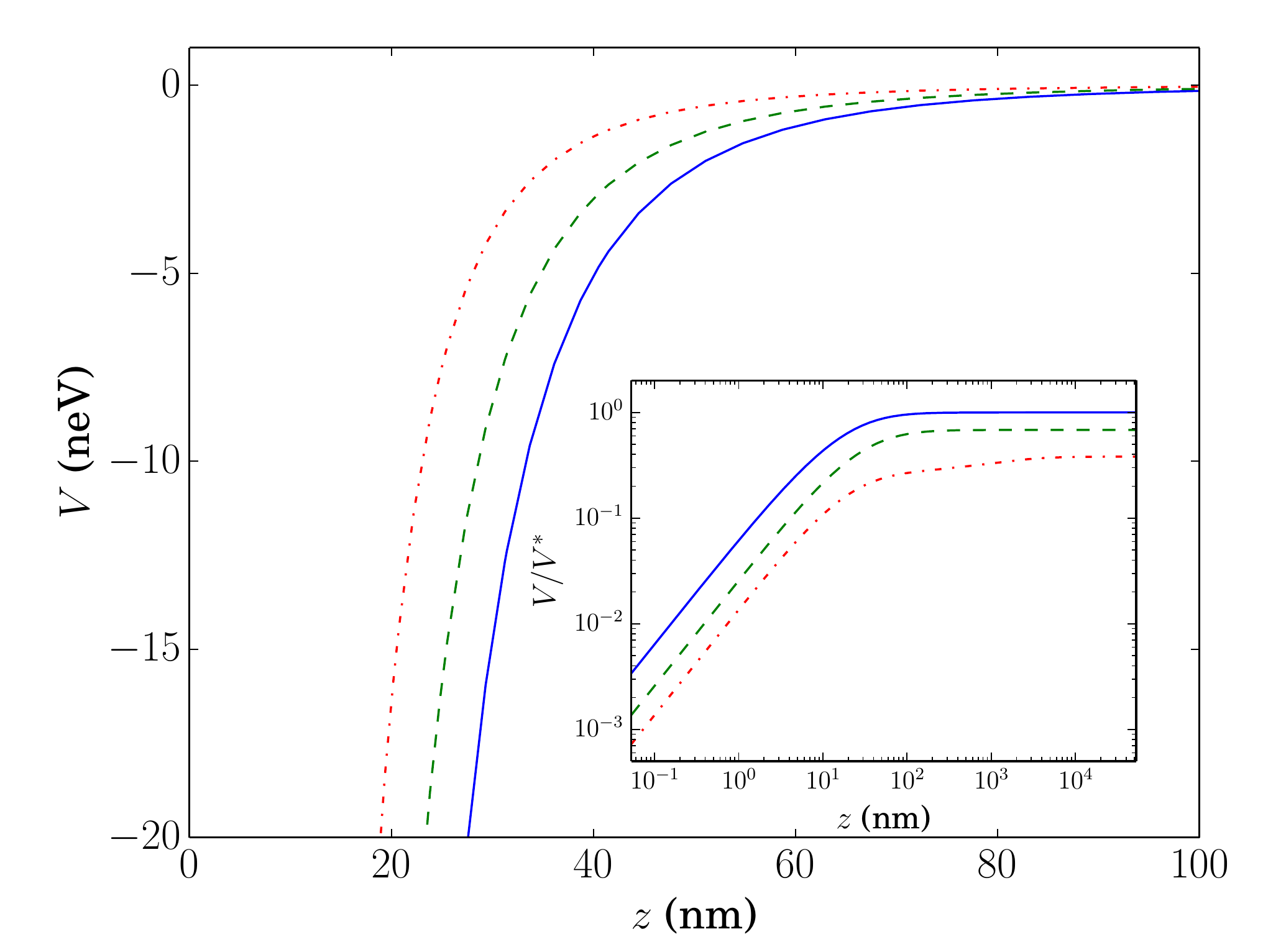}
 \includegraphics[width=0.45\textwidth]{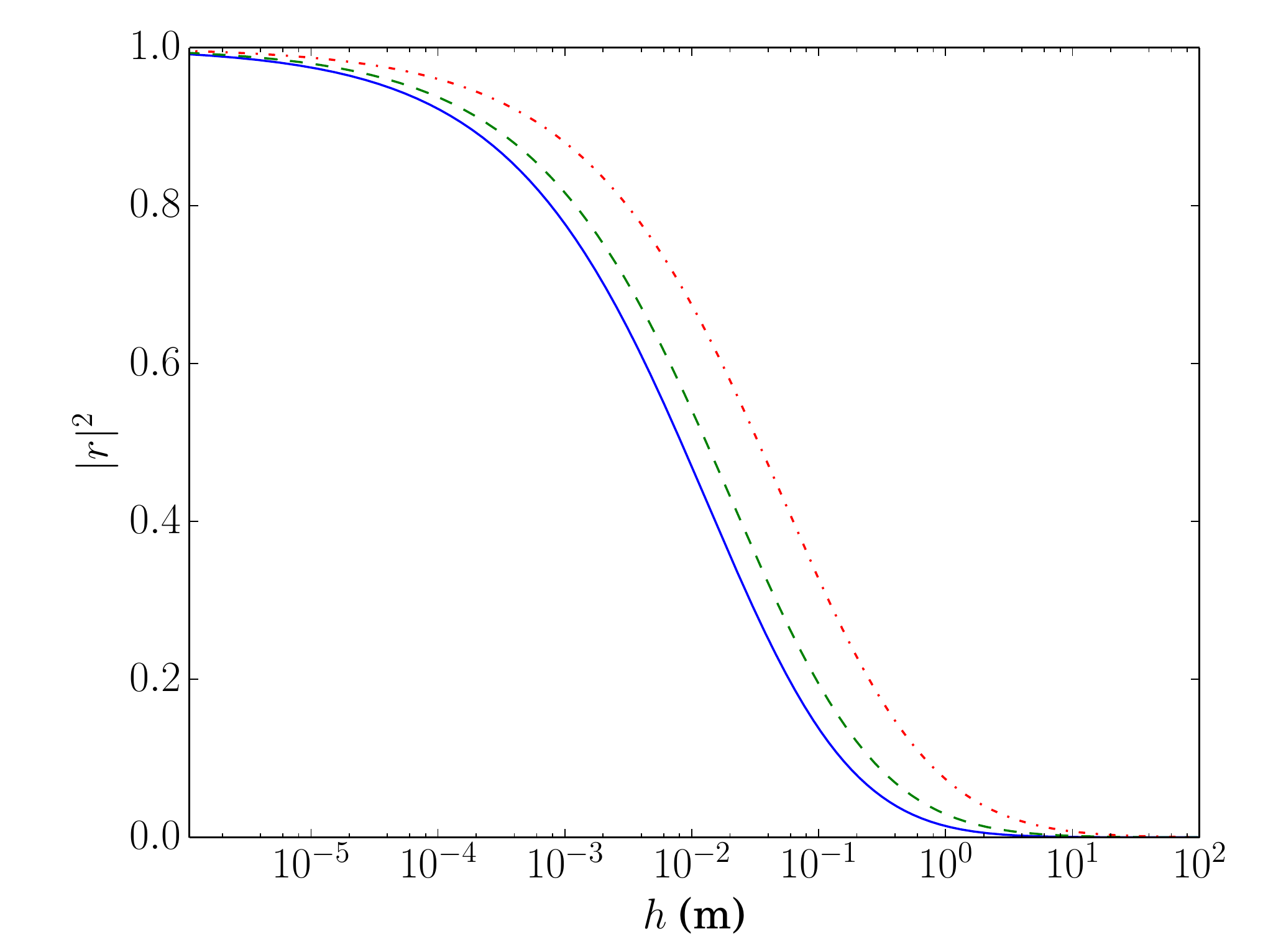}
\caption{\label{potentials-reflectivity} Left panel: Casimir-Polder potential for $\Hb$ near a conducting surface (full blue line), a silicon bulk (dashed green line) and a silica bulk (dash-dot red line). The inset shows the ratio $V(z)/V^*(z)$ to the retarded potential for an ideal mirror (see text). Right panel: Quantum reflection probability for $\Hb$ on the same surfaces (same line styles as the left panel), as a function of the drop height $h=E/mg$.}
\end{figure}

In the left panel of Fig. \ref{potentials-reflectivity} we plot the exact CP potentials between an $\Hb$ atom and a perfectly reflective mirror and 
bulk mirrors made of intrinsic silicon and amorphous silica.
The inset shows the ratios $V(z)/V^*(z)$ to the retarded CP limit calculated for an ideal mirror: $V^*(z)=-C_4^*/z^4$, with
$C_4^*= (3\hbar c/ 8 \pi)(\alpha(0)/4\pi\epsilon_0) = 1.57 \times 10^{-7}$~neV.nm$^4$ for $\Hb$. 
These ratios tend to constant values $C_4/C_4^*\leq1$ at large distances and linear variations $C_3 z/C_4^*$ at small distances. 
The less reflective for the electromagnetic field a material is, the weaker the CP potential, from perfectly reflective to silicon and silica mirrors.

\section{Quantum reflection of $\Hb$ }
\label{sec:qrefl}

The physical problem addressed in this work involves three distinct length scales:
\begin{itemize}
\item the free fall height of $\Hb$ : $h\approx 10$~cm,
\item the scale at which quantization of energy levels in the  gravitational scale becomes important: $\ell_{grav}=\left( \hbar^2/2 m^2 \gb \right)^{1/3} \approx 6$~$\mu$m for $\Hb$, assuming $\gb=g$,
\item the typical range of the Casimir-Polder potential : $\ell_{CP}=\sqrt{2mC_4}/\hbar \approx 30$~nm.
\end{itemize} 
The hierarchy between these lengths scales $\ell_{CP}\ll \ell_{grav} \ll h$ allows to decouple free fall and scattering on the CP potential. 
Therefore, we will solve the Schrödinger equation in the CP potential only, with an energy $E>0$ corresponding to the kinetic energy of the atom before it reaches the CP potential:
\begin{equation}
\label{schrod} \psi''(z) +F(z) \psi(z) =0~, \qquad \qquad F(z)=\frac{2m}{\hbar^2}\left( E-V(z) \right)~.
\end{equation}
Primes denote differentiation with respect to the function's argument. 
To make the connection with the free-fall problem, we will often use the free fall height $h$ as a measure of the energy $E=mgh=102.5$~neV/m $\times h$.

The function $F(z)$  is the square of the de Broglie wave-vector $k_\dB(z)$ associated with the classical momentum $p=\pm\hbar k_\dB$.
Since the potential is attractive, the classical momentum never changes sign: the particle has no classical turning point. In particular, a classical particle moving towards the medium is increasingly accelerated until it hits the surface. 
This classical behavior is mimicked by the WKB wavefunctions which each propagate in a well defined direction:
\begin{align}
\psi_\WKB^\pm(z)= \frac{1}{\sqrt{k_\dB(z)}} e^{\pm i\phi_\dB( z)} ~, 
\qquad  \phi_\dB(z) = \int_{z_0}^z k_\dB(z') \D z'~,
\end{align}
The WKB phase $\phi_\dB(z)$ is proportional to Hamilton's characteristic function associated with the classical trajectory with energy $E$ joining $z_0$ and $z$.
We fix the freedom associated
with the choice of a reference point $z_0$ by enforcing
$\phi_\dB(z)\approx \sqrt{2mE} z/\hbar$ at $z\to\infty$.

In contrast to this semiclassical approximation, the exact wavefunction undergoes quantum reflection. To underline this effect we write the wavefunction in the basis of WKB waves with $z$-dependent coefficients:
\begin{equation}\label{wkbwaves}
\psi(z) = b_+(z) \psi_\WKB^+(z) +b_-(z) \psi_\WKB^-(z)~.
\end{equation}
Introducing this ansatz in Eq. \eqref{schrod} we obtain coupled first-order equations for the amplitudes $b_\pm(z)$, which describe the conversion of an incident wave into a reflected wave \cite{Kemble1935,Berry1972}.

%
In the case of the CP potential, $b_\pm$ go to constant values as $z\to 0$ or $\infty$, so that quantum reflection only occurs in an intermediate  region. The ratio of the amplitudes $b_+(z)$ and $b_-(z)$ as $z$ goes to infinity is the quantum reflection amplitude $r$. 
Because $\Hb$ annihilates if it touches the wall, there can be no outgoing wave immediately above the material surface.
This enforces a full absorption boundary condition $b_+(z=0)=0$.

The right panel of Fig. \ref{potentials-reflectivity} shows the reflection probability $\left\vert r\right\vert^2$ as a function of the energy for each of the potentials calculated in the previous section. 
While the reflection probability vanishes in the classical limit $E\to\infty$, it goes to unity in the purely quantum limit. If not properly accounted for, this energy dependence will bias the detection statistics of GBAR in favor of higher energy atoms.
For an intermediate energy $E=m g \times 10$~cm typical of GBAR the reflection probability is significant: 14\% on a perfectly reflective mirror, 19\% on bulk silicon and 33\% on bulk silica.
Surprisingly, these numbers show that the reflection probability is larger for weaker CP interactions. 
We look into this apparent paradox in more detail in the next section.

\section{Liouville transformations}
\label{sec:liouville}

A Liouville transformation of the Schrödinger equation  \eqref{schrod} consists in a smooth change of coordinate $z \to \tz$ (with $\tz'(z)>0$) and an associated rescaling of the wavefunction: $\tpsi(\tz)=\sqrt{\tz'(z)} \psi(z)$.
Equation \eqref{schrod} for $\psi$ is thereby transformed into an equivalent equation for $\tpsi$ \cite{Olver1997}:
\begin{eqnarray}
\label{transformed}
\tpsi''(\tz)+\tF(\tz)\tpsi(\tz)=0~,\qquad
\tF(\tz)=\frac{F(z)-\frac{1}{2}\{\tz,z\}}{\tz'(z)^2} = z'(\tz)^2
F(z) + \frac{1}{2} \{z,\tz\}~,
\end{eqnarray}
where the curly braces denote the Schwarzian derivative of the coordinate transformation:
\begin{align}\label{schwarzian}
 \{\tz,z\}=\frac{\tz'''(z)}{\tz'(z)}-\frac{3}{2}\frac{\tz''(z)^2}{\tz'(z)^2}~.
\end{align}
Cayley's identity for the Schwarzian derivatives
\begin{eqnarray} \label{cayley}
\left\{\hat z,z\right\} =
\left(\tz'(z)\right)^{2}\,\left\{\hat z,\tz\right\} +
\left\{\tz,z\right\}
\end{eqnarray}
ensures that the composition of two transformations $z\to
\tz$ and $\tz\to\hat z$ is also a transformation $z\to \hat z$.
The inverse transformation, used for writing the last equality in
\eqref{transformed}, is obtained by applying \eqref{cayley} to the
case $\hat z=z$.
The group of Liouville transformations has the remarkable property of preserving the Wronskian of two solutions $\psi_1$, $\psi_2$ of the Schrödinger equation:
\begin{align}\label{wronskian}
 \cW\left(\psi_1,\psi_2\right) =\psi_1(z)\psi_2'(z) - \psi_1'(z)\psi_2(z)=\tpsi_1(\tz)\tpsi_2'(z) - \tpsi_1'(\tz)\tpsi_2(\tz)=\tilde \cW\left(\tpsi_1,\tpsi_2\right)~.
\end{align}
In particular, the reflection and
transmission amplitudes $r$ and $t$ can
be written in terms of Wronskians of solutions which match incoming
and outgoing WKB waves  \cite{Whitton1973}, so they are invariant under the transformation. The probability density
current  $j=\hbar \cW(\psi^*,\psi)/2im$ is also preserved.

We now consider a specific Liouville transformation, where the WKB phase is used as coordinate: $\bz=\phi_\dB(z)$. We use boldfacing to identify quantities related to this coordinate choice. In particular the transformed function \eqref{transformed} can be expressed in terms of a dimensionless energy $\bE=1$ and potential $\bV(\bz)$ given by the ``badlands'' function $Q(z)$:
\begin{gather}
\bF(\bz)=\bE-\bV(\bz)~, \qquad \qquad \bE=1~,\qquad \bV(\bz)=Q(z)~,\\
Q(z)=\frac{1}{2F(z)}\{\phi_\dB,z\}=\frac{F''(z)}{4F(z)^2}-\frac{5F'(z)^2}{16F(z)^3}~.
\end{gather}
In regions where $Q(z)\approx 0$, the WKB wavefunctions are solutions of the Schrödinger equation and there is no reflection. Conversely, the WKB approximation breaks down in regions where $Q(z)$ takes values of order one \cite{Berry1972,Maitra1996,Cote1997,Friedrich2002}, hence the name ``badlands''.

In the case of the CP potential, $Q(z)$ is a peaked function which vanishes far from the surface where the potential goes to zero but also at the surface, where the classical momentum becomes very large. 
The original problem of quantum reflection on a potential well which diverges at one end of the domain $z\in\,]0,\infty[$ is therefore mapped onto an equivalent problem where a particle of unit energy scatters on a potential barrier which vanishes at both ends of the transformed domain $\bz\in\,]-\infty,\infty[$.
The transformed problem is thus a well-defined scattering problem with no
interaction in the asymptotic input and output states. Moreover the transformed problem can have classical turning points where $\bF=0$ or $\bE=\bV$, in which case it corresponds to a radically different semiclassical picture from the original.

\begin{figure}[h]
\centering
\includegraphics[width=0.45\textwidth]{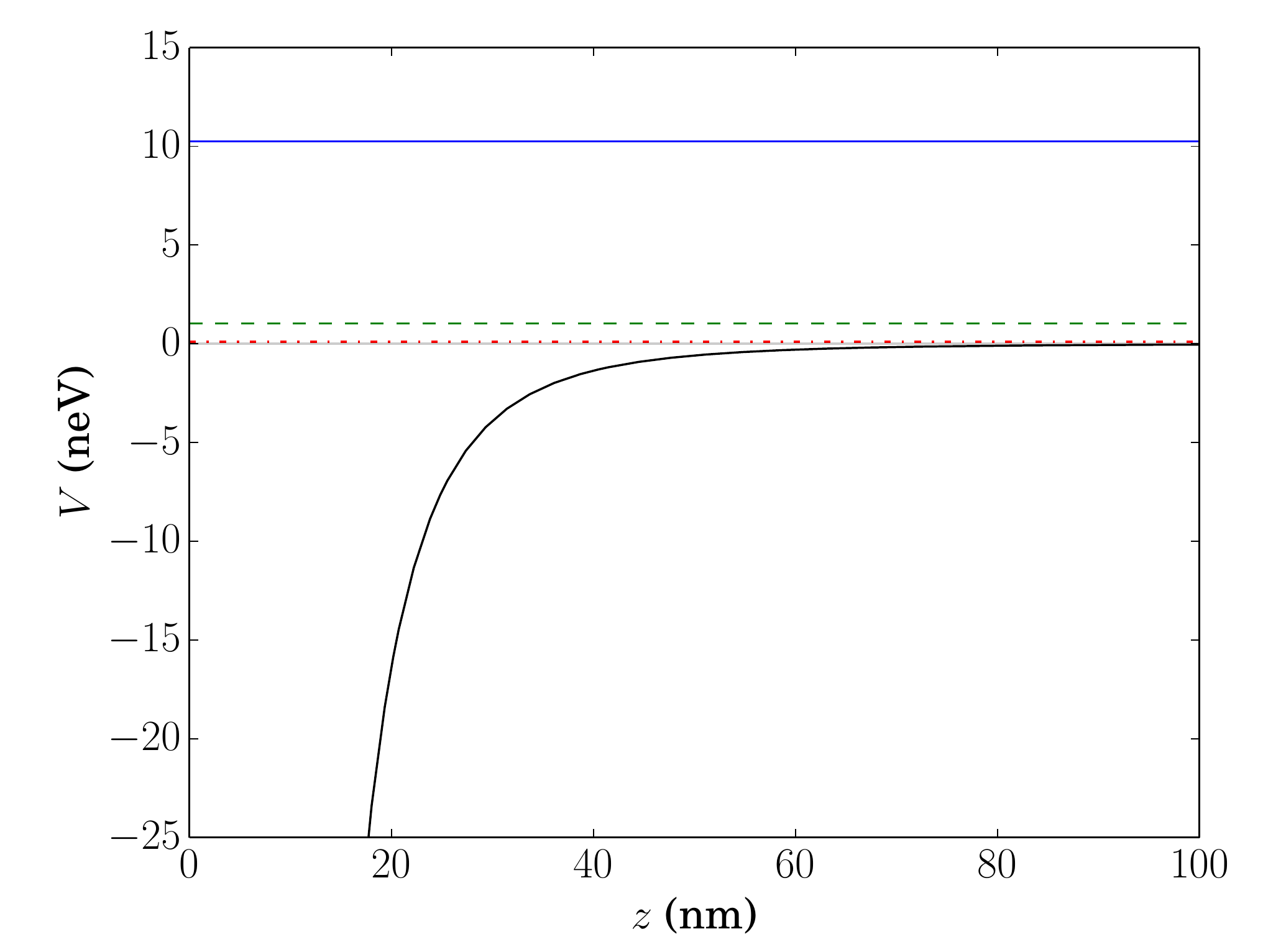}
\includegraphics[width=0.45\textwidth]{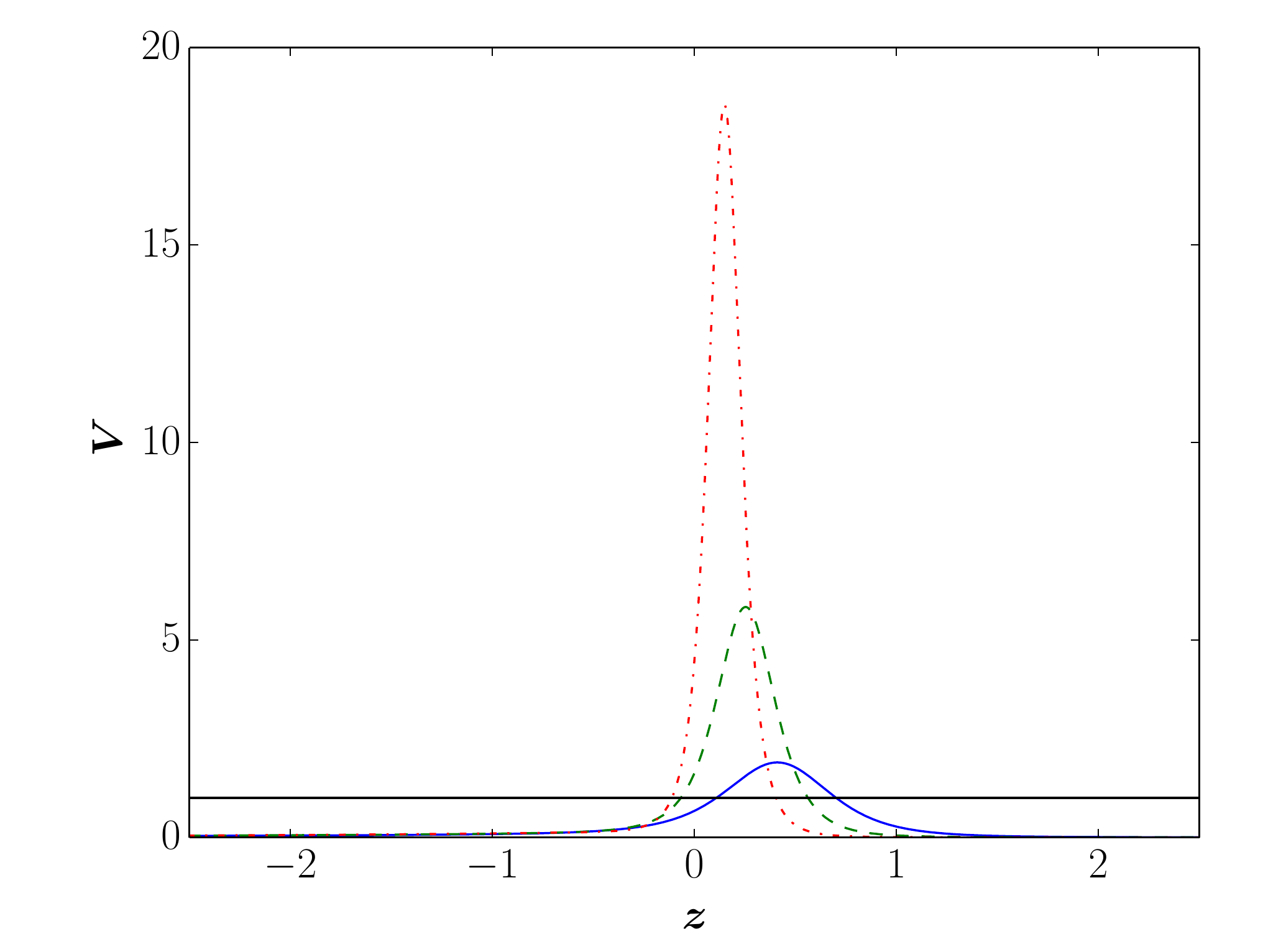}
\caption{\label{energies} Original (left panel) and Liouville-transformed (right panel) energies and potentials for $\Hb$ scattering on a silica bulk with energy $E=m g\times $10~cm (full blue line), 1~cm (dashed green line) and 0.1~cm (dash-dot red line). }
\end{figure}

\begin{figure}[h]
\centering
\includegraphics[width=0.45\textwidth]{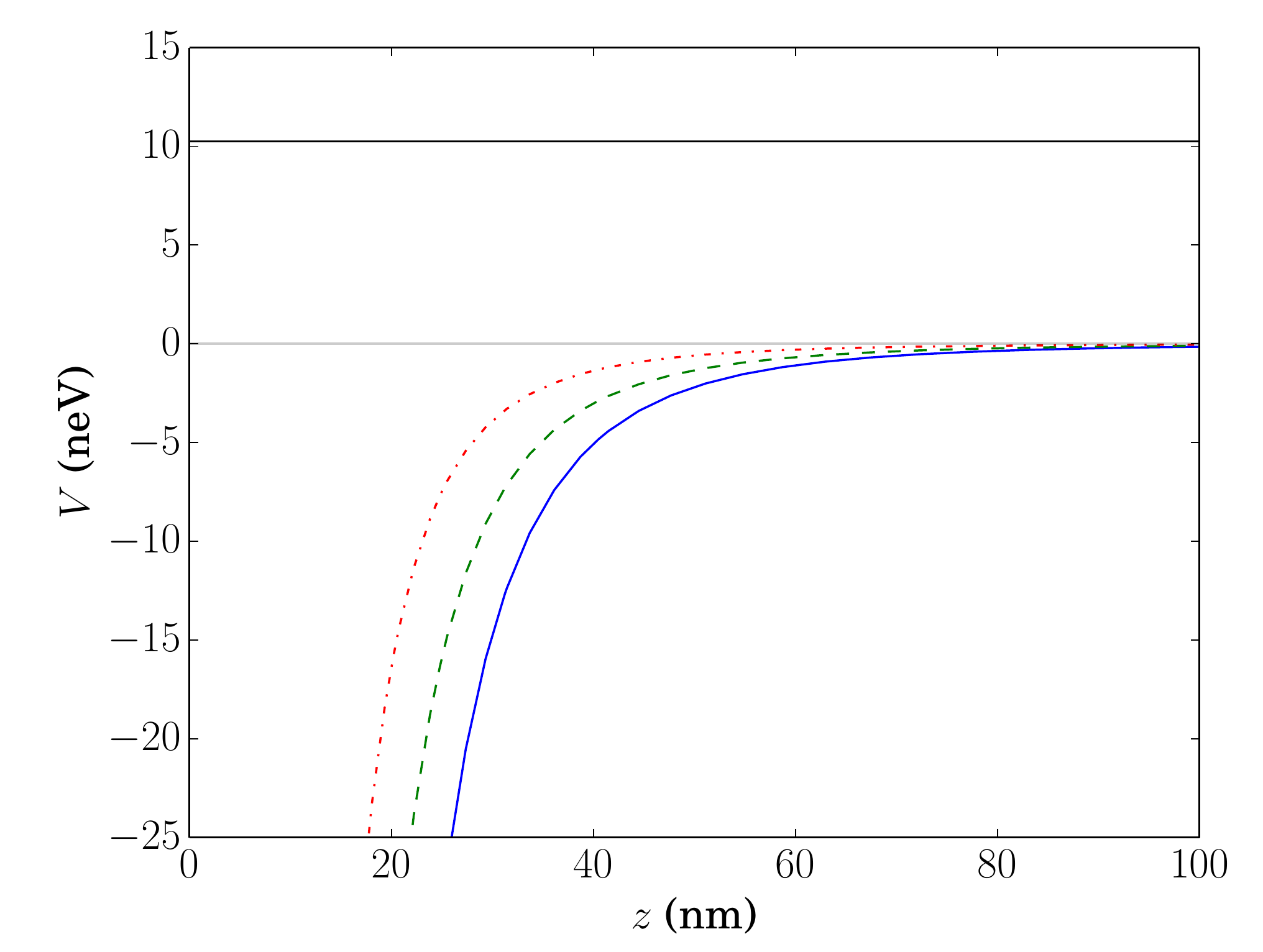}
\includegraphics[width=0.45\textwidth]{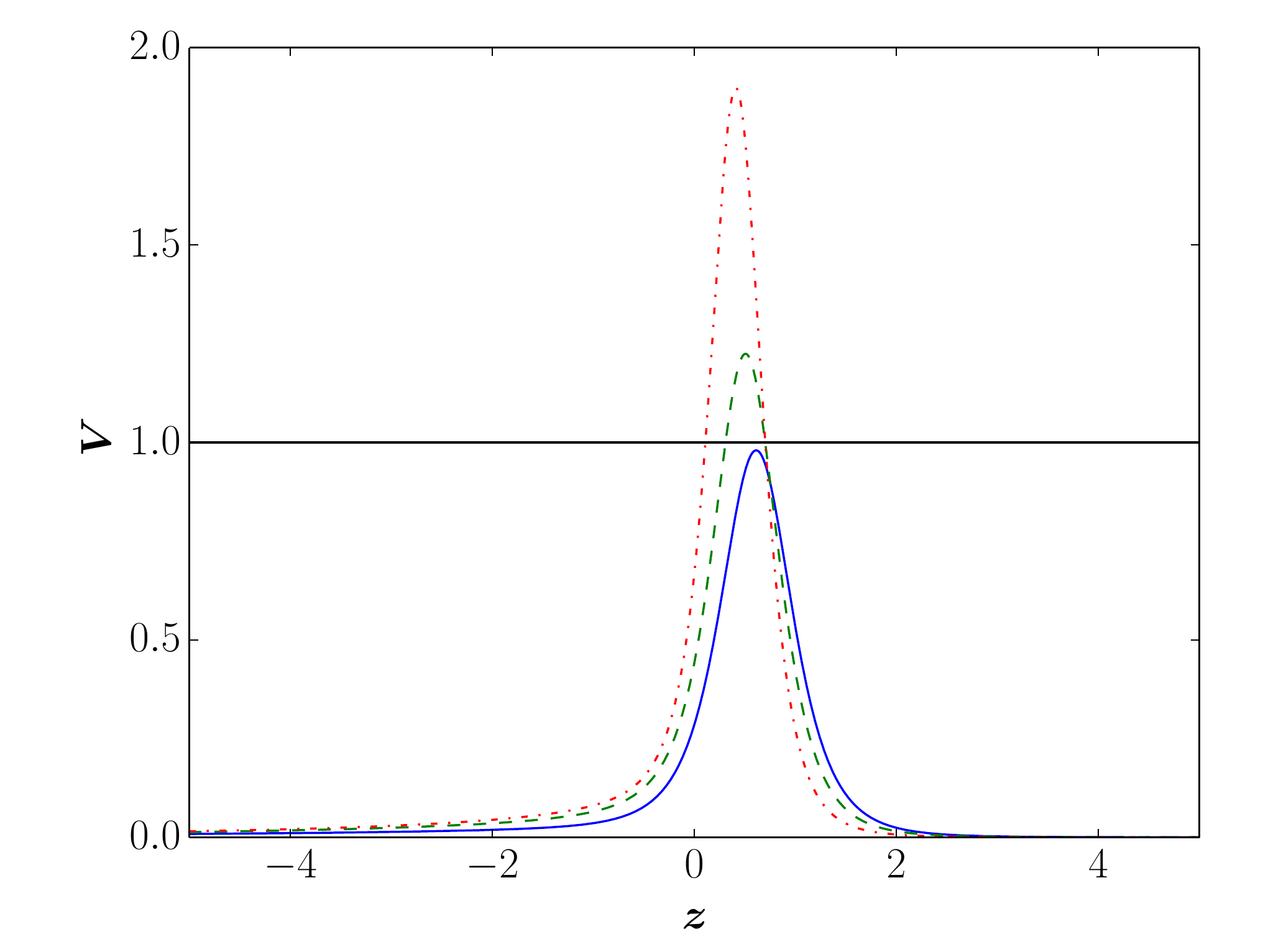}
\caption{\label{mats}  Original (left panel) and Liouville-transformed (right panel) energies and potentials for $\Hb$ impinging with energy $E=m g\times $10~cm on conducting (full blue line), silicon (dashed green line) and silica (dash-dot red line) bulks.  }
\end{figure}

We illustrate this striking fact in Figs. \ref{energies} and \ref{mats} where we show the result of the Liouville transformation for various energies and the potentials presented in section \ref{sec:CP}.
The original quantum reflection problem on an attractive well is now intuitively understood as reflection on a wall, with the same scattering properties.
The height of the barrier grows both when the energy is reduced and when the potential is weakened, which entails a larger reflection probability since the ``energy'' $\bE=1$ is fixed.

\section{Enhancing quantum reflection}
\label{sec:enhanced}

We have seen that a precise evaluation of quantum reflection probabilities is necessary to interpret correctly the results of the GBAR experiment. Beyond this, a good control of quantum reflection opens exciting perspectives for trapping and manipulating antimatter near matter surfaces. For example, the precision of the GBAR experiment is limited by the large initial velocity spread of atoms. This velocity distribution could be tailored  by bouncing the atoms on a mirror above which an absorber is placed to eliminate the faster atoms \cite{Dufour2014shaper}.

In the light of the previous section's results, efficient quantum reflection is obtained for low energies and weak CP interactions. 
At low energies, quantization of the energy levels in the gravitational potential field becomes important. The solutions of the Schrödinger equation in the linear gravitational potential $V(z)=m\gb z$ are expressed with the Airy function:
\begin{align}
\psi(z) \propto \mathrm{Ai}\left( \frac{z}{\ell_{grav}}  - \frac{E}{m\gb \ell_{grav}}\right)~,\qquad \qquad \ell_{grav}=\left( \frac{\hbar^2}{2 m^2 \gb}~. \right)^{1/3}
\end{align}
For an impenetrable mirror the wavefunction vanishes on the surface so the energies $E_n$ of the gravitationally bound states (GBS) are such that $-E_n/m\gb\ell_{grav}$ is a zero of $\mathrm{Ai}$. 
For a real surface, the low energy interaction of $\Hb$ with the surface is described by a complex scattering length $a= \lim_{E\to0} -\hbar\log(-r)/2i\sqrt{2mE}$ \cite{Voronin2005pra,Voronin2005jpb}. This has the effect of displacing the eigen-energies: $E_n \to E_n + m \gb a$ which pick up an imaginary part. The GBS therefore acquire a finite lifetime
$\tau=\hbar/2m \gb |\mathrm{Im}(a)|$ associated with the small probability of $\Hb$ being transmitted through the badlands and annihilated on the surface.

Low CP interaction can be achieved by removing matter from the mirror, so as to reduce its coupling to the electromagnetic field. Nanoporous materials such as silica aerogels incorporate a large fraction of gas or vacuum in 1-10~nm sized pores. For processes involving scales larger than the size of these inhomogeneities, such materials can be described by the Bruggeman effective medium theory \cite{Bruggeman1935,Dufour2013porous}. At the energies of the low-lying GBS, the badlands peak is located $\gtrsim 100$~nm from the surface so $\Hb$ is reflected far enough from the medium for the effective description to be valid.

The lifetimes of GBS above various bulk and porous media are given in Tab. \ref{lifetimes}. In consequence of their weaker CP interaction, the lifetime increases dramatically for porous materials compared with bulk materials, reaching several seconds for silica aerogel. 
Spectroscopy of these long lived GBS could lead to orders of magnitude improvements on the determination of $\gb$  \cite{Voronin2011}.

\begin{table}[h]
\begin{tabular}{|c |c | c | c| c| c| c| } \hline
mirror & perfectly & bulk  & bulk  & silica aerogel  & silica aerogel  & silica aerogel    \\
  &  reflective  &   silicon   &     silica   &(50\% porosity)&(90\% porosity)&(98\% porosity)\\
   \hline
lifetime (s) & 0.11 & 0.14 & 0.22 & 0.32 & 1.07 & 4.64 \\
 \hline
\end{tabular}
\caption{\label{lifetimes} Lifetime in seconds of the first gravitationally bound states of $\Hb$ above various material surfaces.}
\end{table}

\section*{Conclusion}

We have obtained a realistic estimate of the CP potential between $\Hb$ and a variety of material mirrors using the scattering approach to Casimir forces. Quantum reflection on such potentials is the result of a breakdown of the WKB approximation in the so-called badlands region. This is highlighted by performing a Liouville transformation of the Schrödinger equation, which maps quantum reflection to scattering on a barrier given by the badlands function. In this new picture, the increase of the reflection probability when the energy is lowered or the potential weakened is simply interpreted as reflection on a higher peak. Low energy particles can therefore be trapped from above by gravity and from below by quantum reflection. These gravitationally bound states are exceptionally long-lived above nanoporous mirrors and they could be used for highly precise tests of the Equivalence Principle for antimatter.

\section*{Acknowledgments}

The authors wish to thank V. V. Nesvizhevsky, A. Yu. Voronin and the GBAR (gbar.in2p3.fr) and GRANIT collaborations for insightful discussions.

\section*{References}


\end{document}